\documentclass[12pt]{article}
\usepackage{amsmath,amssymb}
\usepackage{graphicx}% Include figure files
\usepackage{caption}
\usepackage{color}% Include colors for document elements
\usepackage{dcolumn}% Align table columns on decimal point
\usepackage{bm}% bold math
\usepackage{float}
\usepackage{hyperref} % hyperref must be loaded before apacite
\usepackage{apacite}
\usepackage{dsfont}
\usepackage[authoryear, square]{natbib}
\hypersetup{ colorlinks = true, citecolor = black, urlcolor = blue}
\usepackage[left=1in, bottom=1in, right=1in, top=1in]{geometry}

\usepackage{eufrak,euscript}
\usepackage{subcaption} %, subfigure}   % <---- DM added subfigure, but there seems to be a conflict, try minipage later??
\usepackage{xr}
\usepackage{ulem}

\newcommand{\argmin}{\operatornamewithlimits{argmin}}
\newcommand{\esssup}{\operatornamewithlimits{ess ~sup}}
\newcommand{\essinf}{\operatornamewithlimits{ess ~inf}}

\title{A Survey of Estimation Methods for Sparse High-dimensional Time Series Models}
\author{Sumanta Basu\thanks{Statistics and Data Science, Cornell University. Email: \href{mailto:sumbose@cornell.edu}{sumbose@cornell.edu}}, David S.\ Matteson\thanks{ILR School and Statistics and Data Science, Cornell University}}
\date{}
\begin{document}
\maketitle

\begin{abstract}
  High-dimensional time series datasets are becoming increasingly common in many areas of biological and social sciences. Some important applications include gene regulatory network reconstruction using time course gene expression data, brain connectivity analysis from neuroimaging data, structural analysis of a large panel of macroeconomic indicators, and studying linkages among financial firms for more robust financial regulation. These applications have led to renewed interest in developing principled statistical methods and theory for estimating large time series models given only a relatively small number of temporally dependent samples. Sparse modeling approaches have gained popularity over the last two decades in statistics and machine learning for their interpretability and predictive accuracy. Although there is a rich literature on several sparsity inducing methods when samples are independent, research on the statistical properties of these methods for estimating time series models is still in progress.

We survey some recent advances in this area, focusing on empirically successful lasso based estimation methods for two canonical multivariate time series models - stochastic regression and vector autoregression. We discuss key technical challenges arising in high-dimensional time series analysis and outline several interesting research directions. 
\end{abstract}

\section{Introduction} 

With rapid advances in information technology, time series datasets in biological and social sciences are becoming high-dimensional, such that one observes a multivariate system with a large number of components over a limited set of time points. This review will highlight several recent methodological and theoretical advances in high-dimensional time series analysis. 

The motivating applications are equally exciting. For examle, in macroeconomic forecasting and structural analyses, researchers often use data from hundreds of series, each containing information on different aspects of an economy \citep{stock2006forecasting}. Key questions in financial risk management revolve around portfolio construction, where one is interested in understanding the covariance matrix of asset returns from hundreds of assets given their historical data \citep{fan2015risks}. Due to structural changes in a constantly evolving economy or financial system, however, the number of stationary observations available for analysis are often of the same order as the number of series. The problem of measuring systemic risk has recently gained considerable interest among policymakers and regulators after the financial crisis of 2007-2009. Empirical measures of connectedness among financial firms are developed by analyzing firm-level characteristics of hundreds of firms to understand { channels of risk propagation}, develop early signals of financial crisis, and identify systemically risky firms \citep{bisias2012survey}.

In biosciences, advances in next-generation sequencing technologies over the past decade, coupled with reduction in experimental costs, are gradually making it possible to measure \textit{genome-wide} expression patterns across thousands of genes and epigenomic elements, at multiple time points and across limited number of replicates (typically of the order of tens) \citep{bar2012studying}. Genomics is expected to produce rich sources of high-dimensional time course datasets in the coming years. In neuroscience, human connectome researchers are interested in finding functional connectivity among hundreds of brain regions using high-dimensional time series datasets (fMRI, MEG, EEG, etc.) \citep{bullmore2011brain}. 

Statistical modeling approaches to address important scientific questions using high-dimensional time series data can be broadly classified into two categories - (i) predictive or forecasting, and (ii) structural estimation. Examples of forecasting topics include predicting the unemployment rates for different sectors of the economy, or identifying firms which are most likely to default at a future time point. Topics related to structure estimation are more common in the biosciences, e.g., understanding regulatory interactions among genes and estimating connectivity among different regions of the human brain. Key questions in economics and finance also require learning how the components of a high-dimensional system interact over time. Some important applications include impulse response function estimation in macroeconomic policy making and identifying financial firms that are ``too-central-to-fail''.

These two classes of problems are not mutually exclusive since consistent structural estimation often helps researchers improve prediction. However, excellent predictions of lower-dimensional components within high-dimensional systems can also be achieved without consistent estimation of the structural properties of the \textit{whole} system. The literature on \textit{estimation of sparse models} for high-dimensional time series, which is the focus of our discussion, is primarily concerned with structural estimation, whereas non-sparse models estimated using shrinkage methods often provide state-of-the-art predictive accuracy in many real-world problems.

With this in mind, we will begin our review with a discussion on stochastic regression, a canonical statistical model in time series, and illustrate the key estimation methods and technical challenges in high-dimension. We will then focus on theory and methods for sparse vector autoregression (VAR), a popular class of systems-identification models, which builds upon stochastic regression techniques and provides insight into interactions among components of multiple time series. For expositional simplicity, we will illustrate the issues of temporal dependence in high-dimension using the popular least absolute shrinkage and selection method (lasso) \citet{tibshirani1996regression}. However, there is a plethora of existing methods, including nonconvex and multi-step procedures, to estimate such models in high-dimension under more general structural constraints. We will allude to some throughout our discussion, but not review their growing adaptations to time series in detail. Estimating factor models will also not be included since we are primarily focused on \textit{sparse} estimation. 

The rest of this article is organized as follows. In section \ref{sec:background}, we start with a short background summary of multivariate time series, and describe different approaches for measuring dependence in these processes, which is crucial for developing meaningful asymptotic theory in high-dimension. Sections \ref{sec:stochreg} and \ref{sec:var} review recent advances in estimation techniques for two classes of high-dimensional time series models - (i) stochastic regression and (ii) vector autoregression (VAR). In Section \ref{sec:var} we also provide information on available software implementations and illustrate a VAR estimation method using an application on a marketing dataset.  We conclude by outlining some interesting research directions in high-dimensional time series analysis. 

\section{Background}\label{sec:background}
\subsection{Stationary Multivariate Time Series}
 
A vector time series is a set of $k$-dimensional vector observations $y_t = (y_{1t},\ldots,y_{kt})'$, each of which is made at a specific time $t$. Here we will focus on real-valued observations, that is, $y_t  \in \mathbb{R}^{k}$, and on discrete-time series, that is, those in which the set of observation times is discrete. We will also suppose observations are made at equally spaced increments. As such, without loss of generality, we will index time by the set of integers, $t \in \mathbb{Z}$.

%\subsection{Examples of Vector Time Series}

%Examples of Vector Time Series
%%
%Show a ?bivariate realization?

%\subsection{Stochastic Vector Processes}

Time series analysis involves selecting a suitable family of models for the observed data. Stochastic models explicitly allow for nondeterministic or random behavior. Such models are natural when future observations may be unpredictable given the past. Here, each observations can be modeled as a realization of a random variable.

A stochastic vector process $\{ y_t \}$ is a sequence of random vector-valued variables.
An observed sequence of data may then be modeled as a single realization of a stochastic process. 
The term time series is often used to refer to both an observed data sequence and a particular stochastic process. 

%$\{ y_t  \in \mathbb{R}^{k} \}$ 

%\subsection{Stationarity}

If $\{ y_t \}$ is a process such that $Var(y_{it}) < \infty$ for each $i \in \{1,\ldots,k\}$ for all $t \in \mathbb{Z}$, then the lagged covariance function $\Gamma(\cdot,\cdot)$ of $\{ y_t \}$ is defined as
$$ \Gamma(r, s) = Cov( y_r, y_s) = E[(y_r - m_r)(y_s - m_s)'], \quad r,s \in \mathbb{Z},$$
in which $m_t = E(y_t)$ for each $t \in \mathbb{Z}$. Then $[\Gamma(r,s)]_{ij} = Cov( y_{ir}, y_{js})$ for all $i,j \in \{1,\ldots,k\}$. It is also important to note that the $k \times k$ matrices $\Gamma(r,s)$ are not symmetric, in general, except for the case in which $r = s$.

A time series $\{ y_t \}$ is stationary if 
(i) $Var(y_{it}) < \infty$ for each $i \in \{1,\ldots,k\}$ for all $t \in \mathbb{Z}$,
(ii) $E(y_t) = \mu$ for all $t \in \mathbb{Z}$,
and (iii) $\Gamma(r,s) = \Gamma(r+t,s+t)$ for each $r,s,t \in \mathbb{Z}$.
These three conditions imply that the mean and lagged covariances are time invariant, that is, they do not depend on the time index $t$. 
This specific definition of stationarity is sometimes referred to as covariance, second-order, or weak stationarity. 
We will take stationarity to mean the above definition, whereas a process $\{ y_t \}$ is called strongly or strictly stationary if $(y_1,\ldots,y_n)$ and $(y_{1+\ell},\ldots,y_{n+\ell})$ have equal joint distributions for all $n \in \mathbb{N}$ and $\ell \in \mathbb{Z}$. A strictly stationary process with finite variances is therefore stationary. 

If $\{ y_t \}$ is stationary then $ \Gamma(r,s) =  \Gamma(0, {s-r})$ for all $r,s \in \mathbb{Z}$, in which case we may redefine the lagged covariance function with respect to only one argument as 
$$ \Gamma(\ell) = \Gamma(0, \ell) = Cov( y_t, y_{t+\ell}) \quad %for \, all \; 
\ell, t \in \mathbb{Z}.$$
The lagged correlation function is similarly defined as $\rho(\ell) = Corr( y_t, y_{t+\ell})$ for all $\ell, t \in \mathbb{Z}$.

%%(Not sure we need to define aCorr fn).
%
%Add more properties under stationarity? as needed?
%
%%The Lagged Covariance Function...of a Stationary Vector Process
%
%%\section{Spectral Representation of a Stationary Vector Process?}
%
%Spectral Representation of a Stationary Vector Process?   The cross spectrum?

\subsection{Measuring Dependence in multivariate time series}
In order to develop efficient algorithms for estimating high-dimensional time series models and study their theoretical properties, it is important to quantify the dependence present in a stochastic process. Datasets generated by processes with stronger temporal dependence are expected to have a smaller effective sample size, and estimates based on these observations converge to model  parameters at a slower rate. %This is because with stronger dependence in the data, the effective sample size becomes smaller and estimators converge to their limits at a slower rate. 

In classical, low-dimensional time series, asymptotic analysis is carried out under the assumption that the dimension ($k$) and model parameters remain unchanged as the sample size ($n$) grows. In contrast, high-dimensional statistics concerns with a so-called \textit{double-asymptotic} regime, in which both the model dimension and the number of parameters are allowed to grow with sample size. 
Theoretical results are typically presented in a \textit{non-asymptotic} format. Here, one derives upper bounds on estimation and prediction errors   (involving $n$, $k$ and other model parameters) that hold with probability converging to $1$, as $k$ grows to infinity \citep{buhlmann2011statistics}. Since this requires controlling finite sample  rather than limiting behavior of time series statistics, more explicit measures of the dependence inherent within a process are needed. For instance, limit theorems in classical time series literature are derived under a short range dependence (SRD) condition $\sum_{\ell=-\infty}^\infty \|\Gamma(\ell)\| < \infty$ \citep{hamilton1994time}. In contrast, developing a suitable non-asymptotic analysis in high-dimension requires explicitly specifying how the partial sums $\sum_{\ell=-n}^n \|\Gamma(\ell)\|$ grow with $n$ and $k$.   

%In the next few sections, 
Before proceeding with our review of stochastic regression and VAR estimation in high-dimension, we provide a short overview of dependence measures used in recent literature on estimation of high-dimensional time series. Some of these measures (e.g., autocorrelation) are easier to apply and interpret for commonly used parametric time series models (e.g., VAR, VARMA, linear processes). Other measures (e.g., mixing conditions) were designed primarily to capture dependence in more general nonparametric time series models, and are typically harder to directly relate to the parameters of simpler linear processes like VARMA. 
% where temporal dependence decay rates are harder to relate to parameters of simpler models. 

% {\color{red} [Add a figure, possibly the convergence rates in high-dimension from the AoS paper. Also, what is a neat notion of stability for stationary processes?]}

\noindent \textbf{Temporal Decay of Autocovariances.} Autocovariance and its scaled counterpart autocorrelation are the most widely used measures of dependence in time series. They capture pairwise and linear measures of dependence only, but are commonly used in the literature because of their simple interpretation. % A covariance matrix can be calculated when considering several random variables, and this idea can be extended to handle infinite sequences of random variables via the autocovariance function $\Gamma(\ell)$. %$\{\Gamma(\ell)\}_{\ell \in \mathbb{Z}}$. 

For specific model classes like first-order Gaussian vector autoregression VAR(1) (see section \ref{sec:var} for details) $x_t = A x_{t-1}+\varepsilon_t$,  a common assumption is $\|A\| < 1$ (\citet{powai2012, hanliu13VAR}). This ensures $\|\Gamma(\ell)\|$ decays geometrically with $\ell$, and hence the sequence $\{\Gamma(\ell)\}_{\ell \in \mathbb{Z}}$ is summable.
%summability of the sequence of autocovariances $\Gamma(\ell)$ in Frobenius norm. 
%{\color{red}  For univariate AR(1) processes, this amounts to assuming the absolute autocorrelation is less than unity, and this assumption holds for all stable AR(1) models.} 
In the special case of scalar AR(1) models ($p=1$), this amounts to assuming the magnitude of the autocorrelation coefficient is less than one, and this assumption holds for all stable AR(1) models.
However, for multivariate processes, the straightforward generalization $\|A\| < 1$ is not satisfied by all stable VAR(1) process. In addition, it does not generalize to higher-order VAR($d$) models, for $d > 1$ \citep{basu2015regularized}.  More generally, for causal linear processes of the form $x_t = \sum_{\ell=0}^\infty A_\ell \varepsilon_{t-\ell}$, $t \in \mathbb{Z}$, in which $\{\varepsilon_t\}$ is an i.i.d.  white noise process (see section \ref{sec:var}), another commonly used criterion is to assume an explicit decay rate on the coefficient matrices $A_\ell$ \citep{chen2016regularized}.

\noindent \textbf{Spectral Density. } \citet{basu2015regularized} 
% analyzed $\ell_1$-regularized and thresholded estimation of a large class of sparse high-dimensional time series models, including stochastic regression, covariance estimation and VAR estimation, with potential extensions to structured sparsity and nonconvex penalized estimates. The authors 
%
considered a class of stable Gaussian time series with bounded spectral density and quantified dependence using a measure of its ``narrowness''. This was inspired by the duality of an autocorrelation function and its Fourier transformation, the spectral density: stronger temporal dependence corresponds to flatter autocorrelation functions, which in turn corresponds to a narrower {or more concentrated} spectral density (Figure  2 in \citet{basu2015regularized}, \citet{Priestley2}). 
%Using this measure of narrowness, the authors derived high probability upper bounds on Lasso penalized stochastic regression and VAR estimates, and thresholding-based covariance matrix estimates proposed in \citep{bickel2008covariance}, and showed that the errors decay at the same rate as i.i.d. data, modulo a ``price'' of dependence which can be captured by the underlying measure of narrowness of the spectral density. 

Formally, for a $k$-dimensional centered stationary Gaussian process $\{x_t\}_{t \in \mathbb{Z}}$, the authors assume that the spectral density $f_x(\theta) = \frac{1}{2\pi} \sum_{\ell=-\infty}^{\infty} \Gamma(\ell) \exp(-i\ell \theta)$ exists and its maximum eigenvalue $\Lambda_{\max}(f_X(\theta))$ is bounded a.e. $\theta$ on $[-\pi, \pi]$, i.e.,  
\begin{equation}\label{defn:measures-stability}
\mathcal{M}(f_x):= \esssup_{\theta \in [-\pi, \pi]} \Lambda_{\max} \left(f_x(\theta) \right) < \infty.
\end{equation}
Such an assumption is satisfied when the autocovariance matrices $\{\Gamma(\ell) \}_{\ell = -\infty}^{\infty}$ are absolutely summable (SRD processes). In particular, this assumption holds for all stable, invertible linear VARMA processes commonly used in classical time series models. 
% Furthermore, it is possible to find tight bounds in terms of ARMA parameters, since the spectral density for this family allows a closed form expression in terms of model parameters. 

$\mathcal{M}(f_x)$ can be viewed as a measure of the ``peak'' of the multivariate spectral density. The authors also define an analogous quantity for the ``trough'' of the spectrum as
\begin{equation*}\label{defn:measures-stability-min}
\EuFrak{m}(f_x):=\essinf_{\theta \in  [-\pi, \pi]} \Lambda_{\min} \left( f_x (\theta)\right).
\end{equation*}

\iffalse
For any subprocess $\{x(J)_t\}_{t \in \mathbb{Z}}$ of $\{x_t\}$, where $J \subseteq \{1, \ldots, k\}$ we can measure dependence using similar quantities. In particular, for any $s \ge 1$, 
\begin{equation*}\label{defn:stability-sub-process}
\mathcal{M}(f_X, s):= \max_{J \subseteq \{1, \ldots, d \}, |J| \le s} \mathcal{M}(f_{X(J)})
\end{equation*}
measures dependence in $s$-dimensional subprocesses. As we will see in next section, this provides insight into how the temporal dependence affects performance of regularized estimates, depending on the complexity of the problem. For instance, for covariance estimation, one only deals with two time series at a time, and the price of dependence scales with $\mathcal{M}(f_x, 2)$. For stochastic regression with an $s$-sparse regression coefficient, the price of dependence scales with $\mathcal{M}(f_x, s)$. 
\fi

Processes with stronger temporal and cross-sectional dependence tend to have larger $\mathcal{M}(f_x)$ and smaller $\EuFrak{m}(f_x)$. Since VARMA processes have a closed form spectral density, upper and lower bounds on $\mathcal{M}(f_x)$ and $\EuFrak{m}(f_x)$ can be derived in terms of the model parameters. This helps interpret how temporal dependence in the observations affects the performance of lasso and other sparse estimation methods. 

Measuring dependence using the narrowness of the spectral density helps develop asymptotic theory for many high-dimensional estimators (e.g., lasso, nuclear norm and nonconvex penalized stochastic regression) on time series generated from stable, invertible VARMA models (\citet{basu2015regularized}, \citet{basu2014modeling}). However, a limitation of this framework is that it relies on Gaussianity, and the results do not directly generalize to nonlinear time series.

\noindent \textbf{Functional/Physical Dependence. } \citet{wu2016performance}, based on the framework proposed in \citet{wu2005nonlinear, wu2011asymptotic} for univariate stationary processes, adopted the notion of functional and predictive dependence measures to quantify dependence in more general nonlinear, non-Gaussian multivariate processes, and established estimation consistency of some regularized estimation procedures for stochastic regression. In earlier work, \citet{chen2013covariance} also adopted this framework to establish consistency for thresholded and $\ell_1$-penalized estimation of high-dimensional covariance and inverse covariance matrices from time series data. 

In this framework of functional and predictive dependence measures, a strongly stationary time series $\{x_t \}$ is assumed to have a data generating process of the form  $x_t = g(\ldots, u_{t-1}, u_t)$, where $\{u_t\}_{t \in \mathbb{Z}}$ are i.i.d.\ random vectors and $g(.)$ is a measurable function. Such a representation, often referred to as a nonlinear Wold representation of stationary processes, has an interpretation in terms of input ($u_t$) and output ($x_t$) sequences of a nonlinear physical system which is denoted by the function $g(.)$. This class of models contains linear stationary  processes, their nonlinear transforms and, under suitable modifications, some locally stationary processes. The \textit{physical dependence} measure is defined as 
\begin{equation}\label{def:physical}
\delta_{t, q} := \|x_t - x^*_t \|_q, \mbox{~~~~ where ~~~} \|x_t \|_q := \left(\mathbb{E}|x_t|^q \right)^{1/q} < \infty, q \ge 1.
\end{equation}
Here, $x^*_t = g(\ldots, u_{-1}, u'_0, u_1, \ldots, u_{t-1}, u_t)$, in which $u'_0$ is an i.i.d.  copy of $u_0$. Each $\delta_{t,q}$ measures how replacing $u_0$ with an i.i.d.  copy changes the output $x_t$. An asymptotic theory is then developed under the short range dependence condition $\Delta_{m,q}:= \sum_{t=m}^\infty \delta_{t, q} < \infty$.

A similar condition of \textit{predictive dependence} is defined as 
\begin{equation*}
\theta_{t,q} = \|\mathbb{E}(x_{t}|\ldots, u_{-2}, u_{-1}, u_0 ) - \mathbb{E}(x_{t}|\ldots, u_{-2}, u_{-1} )\|_q,
\end{equation*}
and asymptotic theory is developed under a short range dependence condition $\Theta_{m,q} = \sum_{t=m}^\infty \theta_{t,q} < \infty$. 
% {\color{red} NEED TO DEFINE $\mathcal{F}_0$ HERE, and why use ell vs t now. Also, typo(s) in line below:}
\citet{wu2016performance} assumed specific decay conditions on $\Delta_{m,q}$ and $\Theta_{m, q}$ as $m \rightarrow \infty$, and established consistency of regularized estimates for high-dimensional time series. These results can be used to quantify dependence over a larger class of processes, but verifying them on stable VARMA processes or causal linear processes requires additional decay assumptions on the model parameters \citep{chen2013covariance}. 

\noindent \textbf{Mixing Conditions. } A more general approach to quantify dependence in high-dimensional time series is to assume specific decay conditions on mixing coefficients used in probability theory to derive limit theorems for dependent random variables. Several mixing conditions exist in the literature, we refer to \citet{bradley2005basic} for an excellent review.  Mixing conditions have been widely used in studying time series and stochastic processes under classical, fixed $p$ asymptotic framework. However, their usage in high-dimensional time series has faced some criticism since they are not easy to verify for many commonly used time series models including VARMA processes. In a recent work, \citet{wong2016regularized} have established consistency of $\ell_1$-regularized VAR estimates with Gaussian and sub-Gaussian innovations under different mixing conditions. We mention two different mixing conditions used in the literature of high-dimensional time series.

For a stationary process $\{x_t\}$, $t \in \mathbb{Z}$, let $\mathcal{F}^s_r = \sigma(x_t, r \le t \le s)$ be the $\sigma$-algebra generated by the random variables $\{x_t \}_{t=r}^s$. For two $\sigma$-algebras of events, $\mathcal{A}$ and $\mathcal{B}$, define 
\begin{equation*}
\alpha(\mathcal{A}, \mathcal{B}) := \sup_{A \in \mathcal{A}, B \in \mathcal{B}} |\mathbb{P}(A \cap B) - \mathbb{P}(A) \mathbb{P}(B)|.
\end{equation*}
Then the process $\{x_t\}$ is called strongly mixing or $\alpha$-mixing if 
$\alpha(m) := \sup_{r \in \mathbb{Z}} \alpha(\mathcal{F}_{-\infty}^r, \mathcal{F}_{r+m}^\infty) \rightarrow 0$ 
as $m \rightarrow \infty$. \citet{rosenblatt1956central} formulated strong mixing condition to establish central limit theorems for dependent random variables. A stronger condition, known as $\beta$-mixing, has also been used earlier in the literature for studying empirical processes and establishing theoretic guarantees with dependent data \citep{vidyasagar2013learning, yu1994rates}. Define 
\begin{equation}
\beta(\mathcal{A}, \mathcal{B}):= \sup \frac{1}{2} \sum_{i=1}^I \sum_{j=1}^J \left| \mathbb{P}(A_i \cap B_j) - \mathbb{P}(A_i) \mathbb{P}(B_j) \right|
\end{equation}
where the supremum is taken over all pairs of (finite) partitions $\{A_1, \ldots, A_I\}$ and $\{B_1, \ldots, B_J\}$ of the sample space such that $A_i \in \mathcal{A}$ for each $i$, and $B_j \in \mathcal{B}$ for each $j$. 
Also, define for each $m \ge 1$, the mixing coefficient 
$\beta(m):= \sup_{r \in \mathbb{Z}} \beta(\mathcal{F}_{-\infty}^r, \mathcal{F}_{r+m}^\infty)$. 
Then the process $\{x_t\}$ is called $\beta$-mixing if $\beta(m) \rightarrow 0$ as $m \rightarrow \infty$. An exponential rate of decay of the $\beta$-mixing coefficients is typically assumed in the literature to ensure meaningful estimation. A technical advantage of $\beta$-mixing, as shown by \citet{yu1994rates}, is that it allows partitioning a collection of dependent random variables into nearly independent blocks. 

Mixing coefficients are routinely used in the probability literature for deriving central limit theorems of sums of dependent random variables. However, their use in developing the concentration inequalities required for high-dimensional time series analysis is still a topic of ongoing research. Despite their analytical amenability and flexibility in accommodating more general forms of nonlinear dependence, mixing conditions have not been widely used in the literature of high-dimensional time series because of their lack of interpretability. For instance, it is often hard to verify mixing conditions and understand how the mixing coefficients $\alpha(m), \, \beta(m)$ change with model parameters in commonly used parametric time series models such as VAR, VARMA, ARCH, GARCH, etc  \citep{wua2011symptotic}. Estimation of these coefficients from data is also not well-studied in the literature. 

\section{Stochastic Regression}\label{sec:stochreg}
Stochastic regression is a canonical problem in classical time series \citep{hamilton1994time}. For a univariate time series of response $\{y_t\}_{t \in \mathbb{Z}}$, a $p$-dimensional vector-valued time series $\{x_t\}_{t \in \mathbb{Z}}$ (possibly including lags of $\{y_t\}$), and a univariate time series $\{\varepsilon_t \}_{t \in \mathbb{Z}}$ of noise, a linear regression model of the form 
\begin{equation}\label{eqn:stochreg}
y_t = \beta_0 + \beta_1 x_{1t} + \ldots + \beta_k x_{kt} + \varepsilon_t, \mbox{~~~} \mathbb{E}(\varepsilon_t | x_t) = 0
\end{equation}
is considered. In our discussion, we will assume $\beta_0 = 0$; all sparsity inducing methods considered below customarily center and scale the response and covariates before fitting a model. After estimating ther regression coefficients $\hat{\beta}_j$, $1 \le j \le p$, intercept is subsequently estimated as $\hat{\beta}_0 = \bar{y} - \sum_{j=1}^p \hat{\beta}_j \bar{x}_j$. %[Some backgrounds from Hamilton?]

\subsection{Estimation}
Based on $n$ realizations $\{(y_t, x_t)\}_{t=1}^n$ from model \eqref{eqn:stochreg}, the traditional technique in low-dimensional setting ($n > p$) is to use ordinary least-squares (OLS) 
\begin{equation*}
\hat{\beta}_{OLS}:= \argmin_{\beta \in \mathbb{R}^p} \|Y - X \beta\|^2,
\end{equation*}
where $Y = [y_1:\ldots:y_n]'$, $X = [x_1:\ldots:x_n]'$ and $\beta = [\beta_1, \ldots, \beta_p]'$ \citep{hamilton1994time}. In a high-dimensional setting $n < p$, the above least-squares problem is ill-posed and OLS is not uniquely defined. Even when $n$ is close to $p$, OLS estimators exhibit high variance and lead to poor predictive performance. In such settings, ridge regression \citep{hoerl1970ridge} has been used as a popular alternative to OLS:
\begin{equation*}
\hat{\beta}_{ridge} = \argmin_{\beta \in \mathbb{R}^p} \|Y - X\beta\|^2 + \lambda \sum_{j=1}^p |\beta_j|^2.
\end{equation*}
Here $\lambda > 0$ is a tuning parameter controlling the amount of shrinkage in the estimation procedure. Ridge induces bias to the estimate, but reduces variance and lowers prediction error in high-dimensional problems for appropriately chosen $\lambda$. However, since ridge does not perform explicit variable selection, all the coefficients in the fitted model are non-zero and the results are not easy to interpret. The lasso \citep{tibshirani1996regression} estimator, defined as  
\begin{equation*}
\hat{\beta}_{lasso} = \argmin_{\beta \in \mathbb{R}^p} \frac{1}{2n}\|Y - X\beta\|^2 + \lambda \sum_{j=1}^p |\beta_j|,
\end{equation*}
presents an attractive alternative by simultaneously shrinking coefficients (hence enabling good prediction) and setting small coefficients to zero, thereby producing a \textit{sparse and interpretable} model. 

Both lasso and ridge have been used for macroeconomic forecasting with high-dimensional time series (see, e.g. \citet{de2008forecasting}) as examples of Bayesian shrinkage. The Bayesian interpretation of these methods is based on the observation that both estimates can be viewed as posterior modes of a Bayesian regression model with i.i.d.  Gaussian (for ridge) or double exponential (for lasso) priors on the regression coefficients.

Other sparsity-inducing penalized estimators have also been used for stochastic regression in recent years. For instance, \citet{medeiros2016l1} considers an adaptive lasso \citep{zou2006adaptive} estimate
\begin{equation*}
\argmin_{\beta \in \mathbb{R}^p} \frac{1}{2n} \|Y - X \beta\|^2 + \lambda \sum_{j=1}^p w_j|\beta_j|.
\end{equation*} 
Here the weights $w_j$, $j=1, \ldots, p$, are chosen as $1/|\hat{\beta}_j|$, with $\hat{\beta}$ being the coefficient of a lasso or elastic net \citep{hastie} model fitted to the data. 
\citet{wu2016performance} considered a Dantzig selector \citep{candes2007dantzig} type estimate
\begin{equation*}
\hat{\beta} = \argmin_{\beta \in \mathbb{R}^p} |\beta_1| \mbox{~~subject to~~} \|X'X\beta - X'Y \|_{\infty} \le \lambda,
\end{equation*}
and study its properties theoretically and using simulation. 

It is well-known in the literature of high-dimensional statistics that these different penalized estimators exhibit different properties in terms of model selection. For instance, lasso tends to select many false positives, while adaptive lasso or a thresholding step after a lasso fit substantially alleviates the problem \citep{van2011adaptive}. In presence of a group of strongly correlated covariates, elastic net guards against false negatives by including all of them in the model, while lasso typically includes only a few and retains a sparse representation. However, in terms of estimation performance, i.e., convergence of $\hat{\beta}$ to the true parameters $\beta^*$, there is no clear case for a single method over others. Asymptotically, all of these estimates show similar rates of convergence. For this reason, we restrict our focus on lasso, which has received a lot of attention in theoretical statistics in recent years.

We conclude this section by pointing to other penalized estimation procedures in the literature of high-dimensional statistics, whose algorithmic and theoretical properties, to the best of our knowledge, have not been explored in the time series context. These include nonconvex penalties like SCAD \citep{fan2001variable}, MCP \citep{zhang2010nearly} that provide better variable selection performance, and more recent proposals of square-root lasso \citep{belloni2011square} and scaled lasso \citep{sun2012scaled} which are easier to tune than lasso using cross-validation. 

\subsection{Theory}
Theoretical properties of lasso in linear regression with i.i.d.  samples have been studied by many authors \citep{knight2000asymptotics, bickel2009simultaneous}. The theory of $\ell_1$-penalized stochastic regression with temporally dependent covariates and errors, however, is still in development. We outline the main technical challenges for extending the existing theory to time series data, and discuss some recent work in this direction. 

Consider a fixed design regression problem $Y = X \beta^* + E$, where $E = [\varepsilon_1: \ldots:\varepsilon_n]'$, and where the parameter $\beta^*$ is assumed to be $s$-sparse, i.e., $\|\beta^*\|_0:= \sum_{j=1}^p \mathds{1}[\beta^*_j \neq 0] \le s$. If $S \subseteq \{1, \ldots, p\}$ denotes the support of $\beta^*$, then  estimation consistency of lasso is established by \citep{powai2012, bickel2009simultaneous} under two assumptions:
\begin{enumerate}
\item \textbf{[Restricted Eigenvalue]: } It is assumed that the null space of the sample Gram matrix $X'X/n$ stays away from an appropriately defined cone set:
\begin{eqnarray*}
\min_{v \in \mathbb{R}^p \backslash \{0\}: \|v\| \le 1,\, v \in \mathcal{C}(S, \alpha)} \frac{1}{n} \|Xv\|^2 \ge \gamma > 0, \mbox{~~ where ~~}  \mathcal{C}(S, \alpha) =    { \{v \in \mathbb{R}^p \, | \, \|v_{S^c}\|_1 \le \alpha  \|v_{S} \|_1 \};}
\end{eqnarray*}
\item \textbf{[Deviation Condition]: } Projection of the error vector on the column space of design matrix is sufficiently small:
\begin{eqnarray*}
\frac{1}{n} \left\|X'E\right\|_\infty \le \varphi(X, \sigma) \sqrt{\log p /n }.
\end{eqnarray*}
Here $\varphi(X, \sigma)$ is a function of the model parameters, and $\sigma = \sqrt{var(\varepsilon_1)}$ is the error standard deviation.
\end{enumerate}
Under the above assumptions, it is shown that the estimation error of lasso with a suitable choice of $\lambda$ satisfies 
\begin{eqnarray*}
\left\|\hat{\beta}-\beta^* \right\| \le \frac{\varphi(x, \sigma)}{\gamma} \sqrt{\frac{s\log p}{n}}.
\end{eqnarray*}
Variants of the above result appeared in the literature under closely related assumptions \citep{buhlmann2011statistics} to ensure estimation consistency of lasso. The first assumption, restricted eigenvalue (RE), is related to the identifiability of $\beta^*$ in sparse linear regression in high-dimension, while the second assumption is required to ensure asymptotic negligibility of correlation between covariates and errors.

In stochastic regression, one works with random realizations of matrices $X$ and $E$, whose rows are not i.i.d.  but consist of consecutive observations from some stochastic process \eqref{eqn:stochreg}. The main technical hurdle in developing a meaningful theory in more general problem lies in understanding how often the two assumptions mentioned above hold under such a data generating process. It is important to note that these conditions are not trivial to verify. For instance, \citet{dobriban2013regularity} shows that verifying RE for a given deterministic matrix $X$ is $NP$-hard. However, when $X$ comes from a random ensemble, it is possible, although highly non-trivial, to show that such an assumption holds with very high probability for a sufficiently large sample size. For instance, if all the entries of $X$ are i.i.d. Gaussian, it is well-known that RE (or even stronger assumptions) holds with high probability when the sample size $n$ is of the order $s \log p$ \citep{mendelson2008uniform}. For correlated Gaussian and subgaussian columns, but independent rows in $X$, similar results appear in \citep{raskutti2010REcorrgauss, rudelsonzhou2012paper}. When the rows are dependent, which is the most common scenario in time series analysis, verifying that these assumptions hold with high probability is challenging. From a technical perspective, this requires developing appropriate concentration inequalities for $S = X'X/n$ around its population mean $\Gamma(0)$, and for $X'E/n$ around zero. Some recent works have made significant progress in this area, as we discuss below.

\citet{basu2015regularized} proved that these assumptions are satisfied with high probability for large enough sample sizes when the underlying processes $\{x_t\}$ and $\{\varepsilon_t\}$ are stationary, Gaussian with bounded spectrum, and are uncorrelated with each other. Under this assumption, \citet{basu2015regularized} derived a non-asymptotic upper bound on the estimation error of lasso for stochastic regression with a suitable choice of $\lambda$. In particular, we have 
\begin{equation*}
\| \hat{\beta} - \beta^* \| \le \frac{\mathcal{M}(f_x) + \mathcal{M}(f_\varepsilon)}{\EuFrak{m}(f_x)} \sqrt{\frac{s\log p}{n}}.
\end{equation*} 
{This bound holds with probability at least $1-c_1 p^{-c_2}$ whenever $n \ge c_3 [\mathcal{M}^2(f_x)/\EuFrak{m}^2(f_x)] s \log p$. Here $c_1, \, c_2$ and $c_3$ are universal positive constants, not depending on the model parameters.} 

\citet{wu2016performance} proved deviation condition (2) under the framework of physical and predictive dependence measures for a large class of short-range dependent processes with subgaussian and heavy-tailed errors. However, their results assume that RE condition holds without verifying it. 

% [ref medeiros] provides a theoretical analysis of adaptive lasso in an asymptotic framework, and allows nonlinearity and heteroskedasticity in the error process. However, the convergence rates seem seem slower when specialized to the case of i.i.d.  data.

\citet{wong2016regularized} have recently shown the validity of above assumptions for linear Gaussian and subgaussian processes under mixing conditions. Interestingly, their convergence rates for subgaussian innovations are slower than the best known rate for Gaussian in \citet{basu2015regularized}. 

At the core of each of the above results is a novel concentration inequality for sums and quadratic forms of dependent random variables. \citet{basu2015regularized} generalized a version of the Hanson-Wright inequality \citep{RV13-hansen-wright} for temporally dependent data generated from stationary Gaussian time series. \citet{wu2016performance} established new results on Nagaev type concentration inequalities for sums of random variables with tails that are possibly heavier than Gaussian. \citet{wong2016regularized} developed a generalized version of Hanson-Wright inequality under $\alpha$ and $\beta$ mixing conditions. These concentration inequalities are potentially useful to investigate other sparsity inducing methods in high-dimensional time series. For instance, \citet{basu2015regularized} show that this techniques can be adapted to conduct non-asymptotic analyses for group sparse, low-rank regression, sparse covariance estimation and nonconvex penalized estimations with high-dimensional time series.

\section{Vector Autoregression}\label{sec:var}
%\subsection{Vector White Noise}

The simplest vector processes are white noise processes. A process $\{ e_t  \in \mathbb{R}^{k} \}$ is a mean zero white noise process with covariance $\Sigma$, denoted $e_t  \sim WN_k(0, \Sigma)$, if and only if $\{ e_t \}$ is stationary, with mean zero, and lagged covariance function $\Gamma(0) = \Sigma$ and $\Gamma(\ell) = 0$ for all $\ell \ne 0$. %for all$\ell \in \mathbb{Z}\backslash 0$
%%Let $e_t  \sim IID_k(0, \Sigma)$ denote an independent and identically distributed process $\{ e_t \}$ with mean zero and covariance $\Sigma$.

%\subsection{Stationary Vector Autoregressive processes} 

A vector process $\{ y_t  \in \mathbb{R}^{k} \}$ is a $d$th order vector autoregressive process, denoted $y_t \sim \text{VAR}_k(d)$ (or simply $\text{VAR}(d)$ when the dimension $k$ is clear from context), if  
\begin{equation}\label{VAR}
y_t = c + \sum_{\ell=1}^d A_{\ell} y_{t-\ell} +  e_{t}, 
\end{equation} 
in which $e_t  \sim WN_k(0, \Sigma)$, $c$ is a length $k$ intercept parameter vector, and each $A_{\ell}$ is a $k \times k$ autoregressive parameter matrix.
The expression above is written more compactly as 
\begin{equation}\label{VARc}
\mathcal{A}_d(L)y_t = c + e_{t}, \quad e_t  \sim WN_k(0, \Sigma),
\end{equation} 
in which $\mathcal{A}_d(z) = I - A_1 z - A_2 z^2 - \cdots - A_d z^d$ is a matrix-valued polynomial in $z$, $I$ is the $k \times k$ identity matrix, and $L$ denotes the Lag or Backshift operator, which is defined by $L^{\ell}y_t = y_{t-\ell}$ for all $\ell \in \mathbb{Z}$.

We will focus our attention on the class of causal and stationary $\text{VAR}_k(d)$ processes. From theorem 11.3.1 of \citep{brockwell1991time}, if 
$$\mathrm{det} [\mathcal{A}_d(z)] \ne 0 \;\; \mathrm{for} \, \mathrm{all} \, z \in \mathbb{C} \; \mathrm{such} \, \mathrm{that} \; |z| \le 1,$$
then (\ref{VAR}) has exactly one stationary solution 
\begin{equation}\label{VARinfty}
y_t = \mu + \sum_{\ell=0}^{\infty} \Psi_{\ell} e_{t-\ell}.
\end{equation}
This solution is a causal function of $\{ e_t \}$, the $\Psi_{\ell}$ are uniquely determined by 
$$\Psi_{\infty}(z) = \sum_{\ell=0}^{\infty} \Psi_{\ell} z^{\ell} = [\mathcal{A}_d(z)]^{-1},$$
with $\Psi_{0} = I$ and $\Psi_{\ell} = \sum_{h = 1}^{\ell} A_h \Psi_{\ell - h}$ for $\ell \in \mathbb{N}$,
the sequence of matrices $\{ \Psi_{\ell} \}_{\ell=0}^{\infty}$ is absolutely summable, and  $\mu = [\mathcal{A}_d(1)]^{-1} c$.
Furthermore, the lagged covariance function of $\{ y_t \}$ is 
$\Gamma_y(\ell) = \sum_{h = 0}^{\infty} \Psi_{h} \Sigma \Psi_{\ell+h}'$ for all $\ell \in \mathbb{Z}$.

A $\text{VAR}_k(d)$ process (\ref{VAR}) may also be represented as a $\text{VAR}_{kd}(1)$ process as follows. 
Let 
$ \tilde y_t = y_t - \mu $, 
$ Y_t = (\tilde y_t', \tilde y_{t-1}', \ldots , \tilde y_{t-d+1}')'$,
and 
$E_t = (\tilde e_t', 0', \ldots , 0')'$
for all $t \in \mathbb{Z}$, such that $Y_t$ and $E_t$ are $(kd \times 1)$,
and let $\tilde A$, the $(kd \times kd)$ companion matrix of the polynomial $\mathcal{A}_d(z)$, and $\Sigma_E$ $(kd \times kd)$ be defined as  
$$ \tilde A = 
\begin{bmatrix} 
A_1 & A_2 & \cdots & A_{d-1} & A_d \\
I & 0 & \cdots & 0 & 0 \\
0 & I & \cdots & 0 & 0 \\
\vdots & \vdots & \ddots & \vdots & \vdots \\
0 & 0 & \cdots & I & 0
\end{bmatrix},
\quad
\Sigma_E = 
\begin{bmatrix} 
\Sigma & 0 & \cdots & 0 \\
0 & 0 & \cdots & 0 \\
\vdots & \vdots & \ddots & \vdots \\
0 & 0 & \cdots & 0
\end{bmatrix},
$$
respectively, in which $I$ is the $(k \times k)$ identity matrix and $0$ is the $(k \times k)$ zero matrix.
Then we have
\begin{equation}\label{VAR1}
Y_t = \tilde A Y_{t-1} + E_{t},
\end{equation}
in which $E_{t} \sim WN_{kd}(0, \Sigma_E)$.
The process $\{ Y_t \}$ is stationary and causal if and only if $\{ y_t \}$ is stationary. Hence, from this $\text{VAR}_{kd}(1)$ representation, we can restate the sufficient conditions with respect to the companion matrix $\tilde A$.
That is, (\ref{VAR1}) has exactly one stationary (and causal) solution if
$$\mathrm{det} [I - \tilde Az] \ne 0 \;\; \mathrm{for} \, \mathrm{all} \, z \in \mathbb{C} \; \mathrm{such} \, \mathrm{that} \; |z| \le 1,$$
in which $I$ is the $(kd \times kd)$ identity matrix.
Furthermore, $\mathrm{det} [I - \tilde Az] \ne 0$ if and only if $\mathrm{det} [\mathcal{A}_d(z)] \ne 0$, \citep{tsay2013multivariate}. 
Finally, $\mathrm{det} [I - \tilde Az] \ne 0$ for all $|z| \le 1$ if and only if $\mathrm{det} [\tilde A - I{z}] \ne 0$ for all $|{z}| \ge 1$,
therefore, we may conclude that an equivalent condition is 
$$ | \tilde{z}_i(\tilde A) | < 1 \;\; \mathrm{for} \, \mathrm{all} \, i \in \{1,\ldots,kd\}, $$
in which $\{ \tilde{z}_i(\tilde A) \}$ denote the eigenvalues of the companion matrix $\tilde A$.

\subsection{Estimation for Vector Autoregression Models} 

Let $\{ y_t \in \mathbb{R}^k \}_{t=1}^n$ denote an observed $k$-dimensional time series with length $n$.
In the low-dimensional setting, the $\text{VAR}_k(d)$ model (\ref{VAR}) can be fit by multivariate least squares.
Here, low-dimensional means not only is $k \ll n$, but $kd \ll n$, as well. 
Let 
$$A = [A_{1},\ldots,A_{d}],$$ 
then the least squares estimate is defined as 
\begin{equation}\label{LS}
\argmin_{c, A} \quad  \sum_{t=1}^n\| y_t - c - \sum_{\ell=1}^d A_{\ell} y_{t-\ell} \|^2. % _2
\end{equation}
%
% in which $\|x\|_2=\sqrt{\sum_{i}x_{i}^2}$ denotes the $L_2$ norm of a vector $x$.
Note that for any $d > 0$ this equation is not well defined unless we also condition on an initial set of $d$ observations $(y_0, y_{-1},\ldots,y_{1-d})'$. These initial values can be treated as missing and imputed (before or during estimation), but that can be quite difficult, in general, and when $n$ is sufficiently large, imputed values should have a negligible influence on the objective function. More commonly, the first $d$ observations are removed and used for initialization, and the $(d+1)$th observation is regarded as the beginning of the time series. The drawback here is that the effective sample size is reduced by $d$ observations.

In general, solving (\ref{LS}) requires estimation of $d(1+kd)$ regression parameters, however, the parameters for each of the $k$ component series may be found by separately minimizing the least squares objective for each of the $k$ marginal (row-wise) expressions as
\begin{equation}\label{LSi}
\argmin_{c_i, A_{i\cdot}} \quad  \sum_{t=1}^n ( y_{it} - c_i - \sum_{\ell=1}^d A_{\ell,i\cdot}' y_{t-\ell} )^2.
\end{equation}
In both approaches, we can see that the number of regression parameters is increasing in both the lag order $d$ and the number of marginal series $k$. 
This illustrates a central tradeoff in VAR modeling: as additional series or lags are included, there is more information to regress on, but there is also greater opportunity to overfit, especially when much of that additional information may be redundant or non-informative. 
However, in many empirical applications, as additional related series are added to the model, a lower lag order $d$ will typically be sufficient in explaining the observed variability.  

VAR estimation in economic applications was popularized by \citet{sims1980}.
Generalized least squares (GLS) is commonly used in regression problems with correlated errors. 
However, the least squares and GLS methods give identical VAR estimates for $c$ and $A$, \citet{zellner1962efficient}.
For Gaussian errors, the maximum likelihood method give asymptotically equivalent estimates. 
Estimates based on the `Yule-Walker' equations are also asymptotically equivalent, but their finite sample bias may be more substantial. 

In the low-dimensional case, selecting the lag order $d$ is commonly performed using an information criterion, such as Akaike's Information Criterion (AIC), its `corrected' version (AICc) or the Bayesian (Schwarz) Information Criterion (BIC), across a limited sequence of $(\tilde d + 1)$ models: $\{ \text{VAR}_k(d) \}_{d=0}^{\tilde d}$. Stepwise chi-square (likelihood ratio) testing and multivariate extensions of the sample partial autocorrelation function are also used for lag identification.

Shrinkage and modern Bayesian VAR models include those by \citet{kadiyala}, \citet{BGR}, \citet{Koop}, and \citet{Gefang}, and early Bayesian approaches include \citet{Litterman1979}, \citet{Litterman1986}, and \citet{Robertson}. 

%A Bayesian implementation of the elastic net is applied for VAR estimation by \citet{Gefang}. The elastic net is a lasso extension proposed by \citet{Hastie}, that is suitable when there is high correlation between covariates.  
%
%Another penalized regression approach is presented by \citet{Gefang} 

Let's now consider the following penalized regression framework for estimating VAR models: 
\begin{equation}\label{VARpen}
\argmin_{c, A} \quad  \sum_{t=1}^n\| y_t - c - \sum_{\ell=1}^d A_{\ell} y_{t-\ell} \|_2^2 + \lambda \mathcal{P}(A),
\end{equation}
in which $\lambda\geq 0$ is a penalty parameter and $\mathcal{P}(A)$ denotes a penalty function on the autoregressive coefficients. Some choices of $\mathcal{P}$ will induce sparsity in estimates of $A$, for suitably large values of $\lambda$.
VAR models with many autoregressive coefficients that are zero may generically be called {Sparse VAR} models.  
For all such methods, careful selection of the penalty parameter $\lambda$ is a crucial component for success. Rolling-validation or cross-validation are typically performed in practice.

Many recent attempts to estimate VAR in high dimensions have utilized the lasso (\citet{tibshirani1996regression}), a least squares variable selection technique. The {`lasso-VAR'} approach was proposed by \citet{hsu} and further considered by \citet{BickelSong}, \citet{li}, and \citet{davis2016sparse}. The lasso-VAR has $\mathcal{P}(A) = \|A\|_1$, in which $\|X\|_1= {\sum_{ij}|X_{ij}|}$ denotes the $L_1$ norm of a matrix $X$.
The {lasso-VAR} has been shown to be more computationally tractable in high dimensions compared with Bayesian methods, and it directly estimates sparse models, as apposed to shrinking parameter estimates towards zero. 

Similar to (\ref{LSi}), when the penalty function $\mathcal{P}(A)$ decouples row-wise as $\{ \mathcal{P}_i(A_{i\cdot}) \}$, we may write (\ref{VARpen}) as
\begin{equation}\label{VARpeni}
\argmin_{c_i, A_{i\cdot}} \quad  \sum_{t=1}^n ( y_{it} - c_i - \sum_{\ell=1}^d A_{\ell,i\cdot}' y_{t-\ell} )^2 + \lambda \mathcal{P}_i(A_{i\cdot}),
\end{equation}
where $A_{i\cdot}$ denotes the $i^{th}$ row of $A$. At this point, the parameters for each full marginal model can again be estimated independently. However, selection of $\lambda$ still requires the results from all $k$ components. When the error variances may be unequal across the component models, it may be useful to select a separate penalty parameter $\lambda_i$ for each of the $k$ regressions. However, the situation is more complicated when the errors are also contemporaneously dependent.

%\citet{tsay2013multivariate}

Lag structured penalty functions were recently proposed by \citet{Nicholson2017627}. These {`VAR-L'} models are sparse VAR models in which the sparsity pattern is directly linked to the coefficients' lag order. There is also an explicit distinction between a 
component series' own lags and the lags of other series. The authors also derived efficient estimation algorithms for this flexible class of sparse VAR models. 

%Let $\|X\|_F=\sqrt{\sum_{ij}X_{ij}^2}$ denotes the Frobenius norm of a matrix $X$, and
%let $X^{\text{on}} = I \circ X$ and $X^{\text{off}} = X - X^{\text{on}}$, in which $\circ$ denotes the elementwise product. 

The simplest lag-based penalty function within the VAR-L framework is the `lag group' penalty, which is defined as  
\begin{equation}\label{laggroup}
\mathcal{P}(A) = \sum_{\ell=1}^d\| A_{\ell}\|_F,  %\sqrt{d^2}
\end{equation}
in which $\|X\|_F=\sqrt{\sum_{ij}X_{ij}^2}$ denotes the Frobenius norm of a matrix $X$.
This is simply the sum of the Frobenius norm of each lagged autoregressive coefficient matrix $A_{\ell}$. 
This penalty structure implies that an estimate of $A_{\ell}$ will be shrunk towards zero, but its elements will either all be zero or all non-zero, for each lag $\ell = 1,\ldots,d$. That is, the sparsity pattern in estimates of $A$ will exhibit group sparsity, in which the groups are defined with respect to the coefficients' lag order $\ell$.

In some applications, the lag group penalty may be too restrictive. However, this can easily be extended to allow for within group sparsity using the `sparse lag group' VAR-L penalty, which is defined as  
\begin{equation}\label{sparselaggroup}
\mathcal{P}(A) = (1-\alpha)\sum_{\ell=1}^d\| A_{\ell}\|_F+\alpha\| A\|_1,
\end{equation}
for $\alpha \in [0,1]$. Note that for $\alpha = 0$ this is the `lag group' penalty function, and for $\alpha = 1$ this is the $L_1$ norm, i.e., the penalty function used in the lasso-VAR approach, or the `basic' VAR-L penalty function. 
In general, $\alpha$ balances the group and within group sparsity levels for estimates of $A$. 
To balance the relative size of the lag groups with the singleton groups implied by the $L_1$ norm, letting $\alpha = \frac{1}{k+1}$ was suggested. However, it may also be selected through rolling- or cross-validation, along with the penalty parameter $\lambda$. 

The sparse lag group structure is very flexible, but it gives equal consideration to all elements within a lagged coefficient matrix. However, in many applications, it is the diagonal elements of $A_{\ell}$ that are most likely to be non-zero. These diagonal elements correspond to each marginal series being regressed on its `own' lagged values, while the off-diagonal coefficients correspond to regression on the lagged values of the `other' components. 
To allow separate treatment of these different types of predictors, the VAR-L framework includes the `own/other' and the more general `sparse own/other' lag group penalty functions. 

Let $X^{\text{on}} = I \circ X$ and $X^{\text{off}} = X - X^{\text{on}}$ for a matrix $X$, in which $\circ$ denotes the elementwise product. Then, the `sparse own/other' lag group penalty function is defined as 
\begin{equation}\label{sparseOOlaggroup}
\mathcal{P}(A) = (1-\alpha)\big(\sqrt{k}\sum_{\ell=1}^d||A^{\text{on}}_{\ell}||_F+\sqrt{k(k-1)}\sum_{\ell=1}^d||A^{\text{off}}_{\ell}||_F\big)+\alpha\| A\|_1
\end{equation}
for $\alpha \in [0,1]$. Similar to above, this is simply the `basic' VAR-L or lasso-VAR penalty function for $\alpha = 1$, the `own/other' lag group VAR-L penalty function for $\alpha = 0$, and $\alpha = \frac{1}{k+1}$ was again suggested, but it may also be selected. 

By extending the VAR model to also include regression on (lagged) exogenous variables (often called a VARX model), the VAR-L methods are imbedded within the larger VARX-L framework, as detailed in \citet{Nicholson2017627}. 
Two additional methods are proposed within this framework. Shrinkage towards a given model, or a given set of model parameter values, and in particular, shrinkage towards a vector random walk model via the `Minnesota' VARX-L penalty function. An endogenous-first active set using a hierarchical penalty function was also proposed. It allows non-zero exogenous variable coefficient estimates only if the corresponding endogenous variable coefficients are also estimated to be non-zero, at a given lag.

The lasso-VAR or basic VAR-L penalty functions penalize all lagged coefficients equally. The various lag group VAR-L penalty function similarly apply the same penalty at each lag. 
A {lag-weighted lasso-VAR} is proposed in \citet{BickelSong}, in which the 
penalty function is a weighted $L_1$ norm, with weights that increase with the lag order:
\begin{equation}\label{lagweighted}
\mathcal{P}(A) = \sum_{\ell=1}^d\ell^\gamma\|A_{\ell}\|_1,
\end{equation}
for $\gamma \in [0,1]$.
The parameter $\gamma$ controls the rate at which the weights increase with respect to the lag.
Rolling-validation was suggested for selecting this additional tuning parameter.

Applying a greater penalty to coefficients with higher lag orders leads to estimates with greater sparsity as the lag increases. 
However, the weighted approach will not select a maximal lag, beyond which all later coefficients are estimates as zero. 
The hierarchical lag-based penalty functions proposed in \citet{nicholson2014hierarchical} embeds the notion of lag selection for VAR models into a convex regularizer, with computationally efficient algorithms that can be parallelized across the component models. 
The motivation for these `HVAR' models is very simple, short-term lag-lead relationships are much more likely to exist than very long-term lag-lead relationships, in many applications. 
An additional benefit of the HVAR methods is a highly interpretable maxlag matrix, which can be visualized with a heat-map that denotes the persistence of all pair-wise lag-lead VAR relationships, even in high dimensions.

The HVAR framework includes three hierarchical lag-based penalty structures: componentwise, elementwise, and own/other. 
For $1\le\ell\le d$, let
\begin{align*}
%  {A}^{(\ell:d)}&=[A^{(\ell)}~\cdots~A^{(d)}]\in\mathbb
 % R^{k\times k(d-\ell+1)}\\
  {A}_{i}^{(\ell:d)} =[A_{\ell,i\cdot}'~\cdots~A_{d,i\cdot}']
  \in\mathbb R^{1\times k(d-\ell+1)}   \quad \quad and  \quad \quad
  {A}_{ij}^{(\ell:d)} =[A_{\ell,ij}~\cdots~A_{d,ij}]\in\mathbb
  R^{1\times (d-\ell+1)}.
\end{align*}
The {`componentwise'} hierarchical lag penalty function for {HVAR$^{C}$} is defined as
\begin{align}
\label{OV1}
\mathcal{P}(A) = \sum_{i=1}^k\sum_{\ell=1}^d\|{A}_i^{(\ell:d)}\|_F.
\end{align}
This implies a componentwise hierarchical lag group structure, and imposes the following (row-wise) active set condition on estimates $\hat A$:
if $\hat{{A}}_{\ell, i \cdot}=0$, then $\hat{{A}}_{h, i \cdot}=0$ for all $h>\ell$, for each $i = 1,\ldots,k$. 
The {`elementwise'} hierarchical lag penalty function for {HVAR$^{E}$} is similarly defined as 
\begin{align}
\label{OV3}
\mathcal{P}(A) = \sum_{i=1}^k\sum_{j=1}^k\sum_{\ell=1}^d\|{A}_{ij}^{(\ell:d)}\|_F.
\end{align}
This implies a hierarchical lag group structure for each of the $k^2$ pairs of components. It imposes the following active set condition on estimates $\hat A$: 
if $\hat{{A}}_{\ell, i j}=0$, then $\hat{{A}}_{h, i j}=0$ for all $h>\ell$, for each $i,j = 1,\ldots,k$. 
While the {HVAR$^{E}$} approach is more flexible, and all pairs of components can have different lag-truncated sparsity levels,
it also does not directly share information across the different components as the componentwise approach does. 

Finally, the {`own-other'} hierarchical lag penalty function for {HVAR$^{O}$} was inspired by the own/other lag group VAR-L penalty function. It is defined by
 \begin{align}
\mathcal{P}(A) = \sum_{i=1}^k\sum_{\ell=1}^d\left[\|{A}_{i}^{(\ell:d)}\|_F+\|A^{\text{off}}_{\ell}, A_{i}^{([\ell+1]:d)})\|_F\right],
 \label{OV2}
 \end{align}
in which 
%$A_{i,-i}^{(\ell)}=vec\{A_{\ell, ij}:j\neq i\}$,
%$A_{i,-i}^{(\ell)}= A^{\text{off}}_{\ell}$, 
%and where 
we define $A_{i}^{([d+1]:d)}={0}$.
The first part of this penalty function matches \eqref{OV1}, and the second part is the same as the first, except the diagonal elements of $A_{\ell}$ are not included.  
This hierarchical structure imposes the following active set condition on estimates $\hat A$: 
for all $h>\ell$,
if $\hat{{A}}_{\ell, i\cdot}={0}$ then $\hat{{A}}_{h,i\cdot}={0}$, and if $\hat{{A}}_{\ell,ii}={0}$ then $\hat{{A}}_{\ell,ij}={0}$ for all $j$. %$\hat{{A}}_{h+1}^{\text{off}}={0}$.
This means that as the componentwise lag group order increases, a series own lagged coefficient must become active before those of the other series, at that lag. 
 
% The HVAR framework has also been extended to include regression on (lagged) exogenous variables (HVARX), see \citet{wilms2017HVARX}. 

\subsection{Theory}
Theoretical properties of penalized VAR estimates have been studied by several authors in recent years. For the sake of brevity, we focus only on Lasso penalized VAR. Additional works on adaptive lasso, group penalized VAR and low-rank estimation of VAR models include \citet{medeiros2016l1, BickelSong, negwai2011}.  Since VAR estimation can be posed as $k$ separate regularized regression problems, theoretical challenges due to temporal and cross-sectional dependence in data are similar to the ones discussed in Section \ref{sec:stochreg}. 

\citet{BickelSong} established non-asymptotic upper bounds on lasso and group lasso penalized VAR estimates under a fractional cover based dependence condition on the underlying processes. However, it was noted that even for VAR(1) processes, the bounds could be arbitrarily large. \citet{hanliu13VAR} established non-asymptotic upper bound on estimation and in-sample prediction errors of lasso and Dantzig selector type estimates of sparse, high-dimensional VAR(1) models. The authors show that suitable restricted eigenvalue and deviation conditions are satisfied with high probability, when data comes from a VAR model and the sample size is sufficiently large. The authors assume that the spectral norm of transition matrix is smaller than unity ($\|A\| < 1$). \citet{basu2015regularized} studied stable Gaussian VAR(d) models with $d \ge 1$ and established non-asymptotic upper bound on the estimation of $\ell_1$-regularized least squares and log-likelihood based estimates under the spectral density based dependence conditions. \citet{wu2016performance} established estimation error bounds assuming RE conditions for a larger class of non-Gaussian VAR models with possibly heavy-tailed distributions, under the dependence framework of functional dependence. A key technical ingredient of their work is new Nagaev type concentration inequalities for sums of dependent variables. 

More recently, \citet{wong2016regularized} established upper bounds of Lasso penalized VAR estimates under mixing conditions. In particular, they verify necessary RE and deviation conditions for sub-Gaussian innovations under $\beta$-mixing conditions on the underlying data generating process, and for Gaussian VARs under a weaker $\alpha$-mixing condition. Some recent work on heavy tailed VAR models can be found in \citet{qiu2015robust, wong2017lasso}. \citet{medeiros2016l1} recently studied properties of adaptive lasso estimates under mixingale type assumptions \citep{davidson1994stochastic} commonly used in econometric literature. Their results are asymptotic in nature, and we do not provide a comparison with the non-asymptotic results described above. 

The above results seem to follow a central theme observed in the case of stochastic regression as well. With high probability, the estimation error is upper bounded by the same rate as in i.i.d.\ sample, but scaled by a ``price'' due to temporal and cross-sectional dependence within the observations. Depending on the dependence framework used in the analysis, this ``price'' could be a function of $1 - \|A\|$ \citep{powai2012, hanliu13VAR}, narrowness of underlying spectra \citep{basu2015regularized}, physical/functional dependence \citep{wu2011asymptotic}, or the rate at which the mixing coefficients decay \citep{wong2016regularized}. The results on group sparse and low-rank estimation of VAR also seem to fit this criterion \citep{negwai2011, basu2014modeling}. In each case, this provides a theoretical justification for using modern penalized and constrained estimation procedures to fit sparse models to high-dimensional time series under stationarity and short-range dependence assumptions. However, several interesting questions remain regarding the extension of sparsity-based methods to nonstationary and long-memory VARs in high-dimension.

\subsection{Software and an Illustrative Example}

There are several existing R packages for regularized estimation, but the BigVAR package for R \citep{BigVAR, nicholson2017bigvar} provides computationally efficient implementations of the lasso-VAR and weighted lasso-VAR models, as well as all models in the VAR-L, VARX-L, and HVAR frameworks.
The bigtime package for R \citep{bigtime} also includes the HVAR framework, as well as extensions with (lagged) exogenous variables \citep{wilms2017interpretable}, and VARMA modeling \citep{wilms2021sparse}.

\begin{figure}
    \centering
    \begin{subfigure}[b]{0.3\textwidth}
        \includegraphics[width=\textwidth]{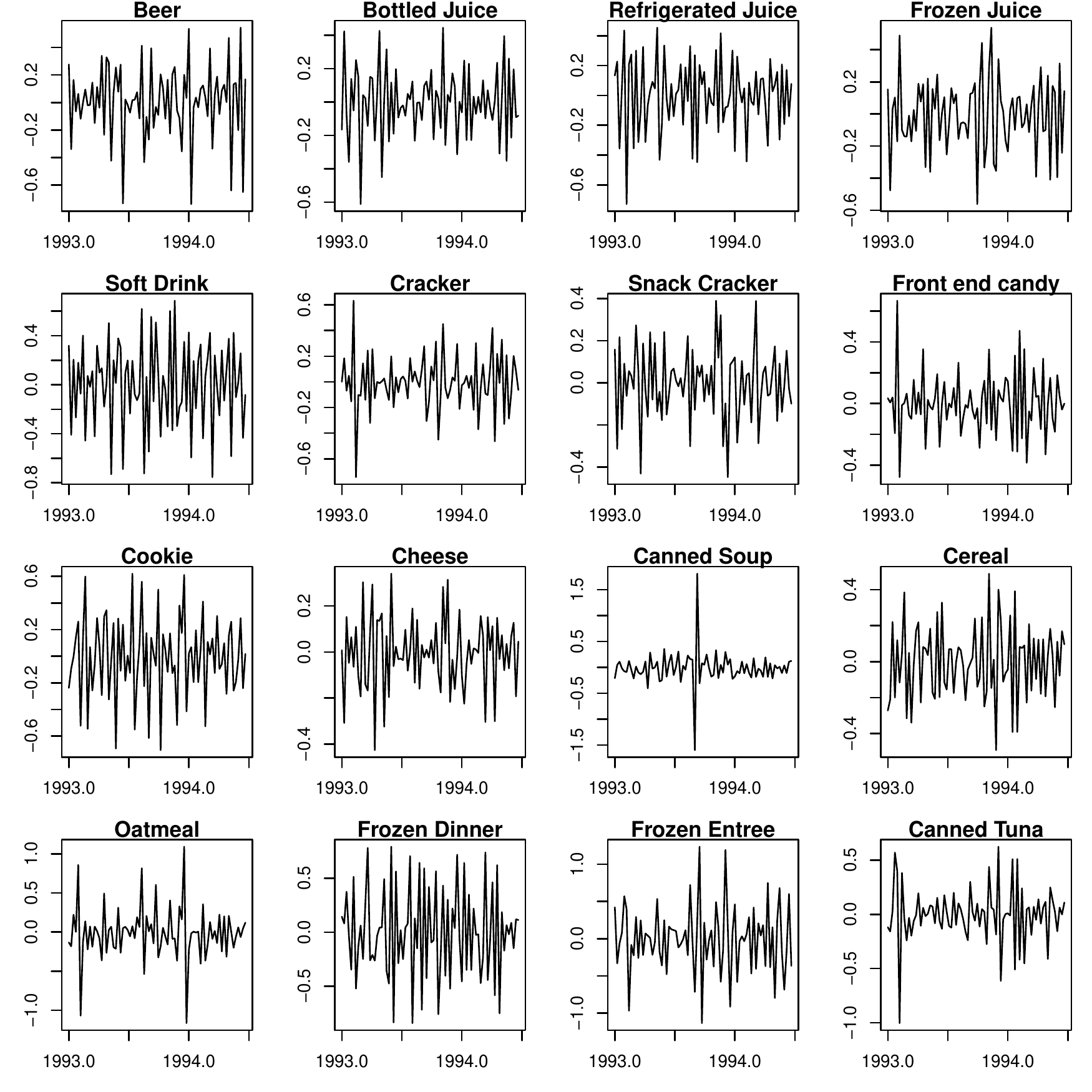}
        \caption{Monthly Sales Growth} % Time series of monthly sales growth for the categories
        \label{fig:3waya}
    \end{subfigure}
    ~ %add desired spacing between images, e. g. ~, \quad, \qquad, \hfill etc. 
    \begin{subfigure}[b]{0.3\textwidth}
        \includegraphics[width=\textwidth]{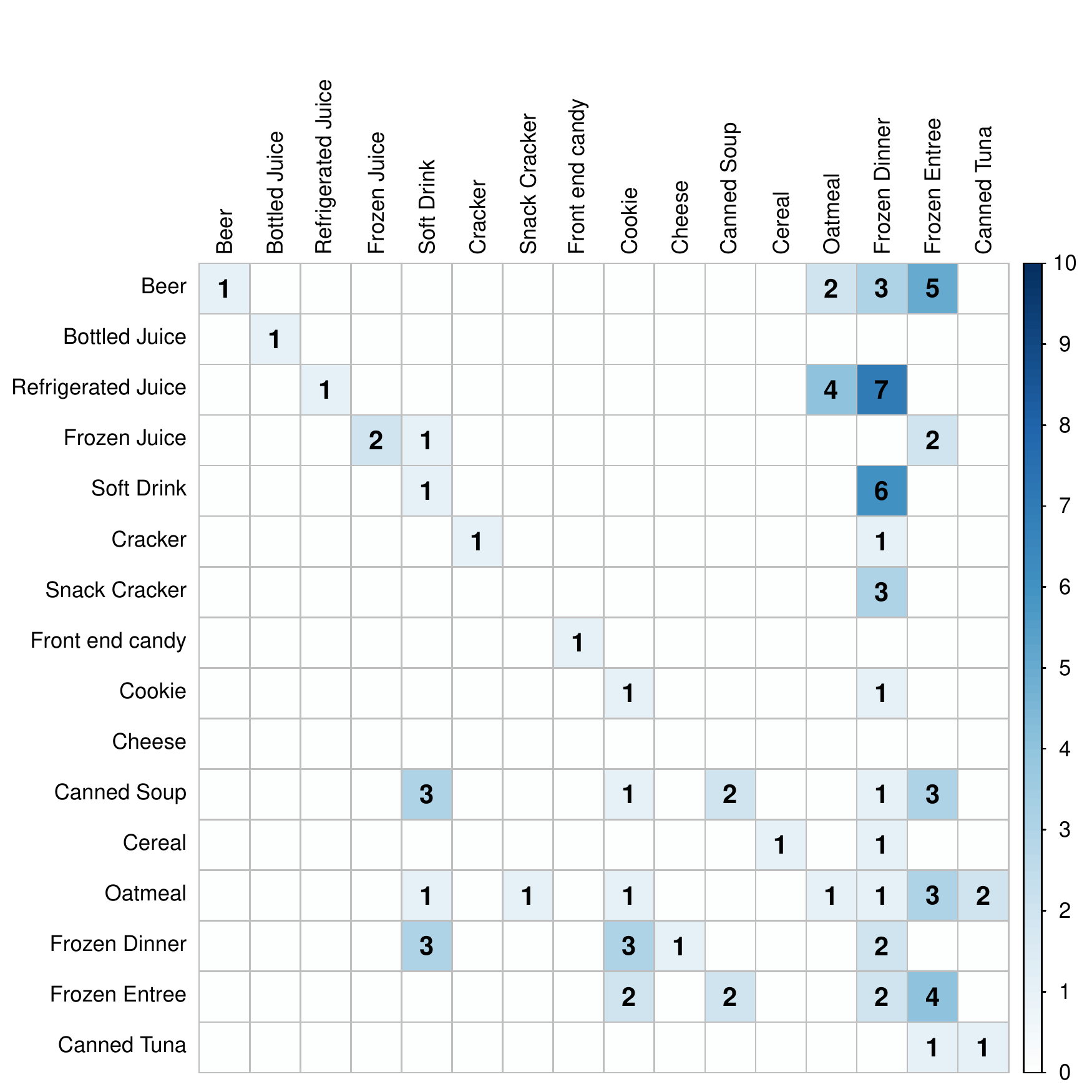}
        \caption{Lag matrix}
        \label{fig:3wayb}
    \end{subfigure}
    ~ %add desired spacing between images, e. g. ~, \quad, \qquad, \hfill etc. 
    \begin{subfigure}[b]{0.3\textwidth}
        \includegraphics[width=\textwidth]{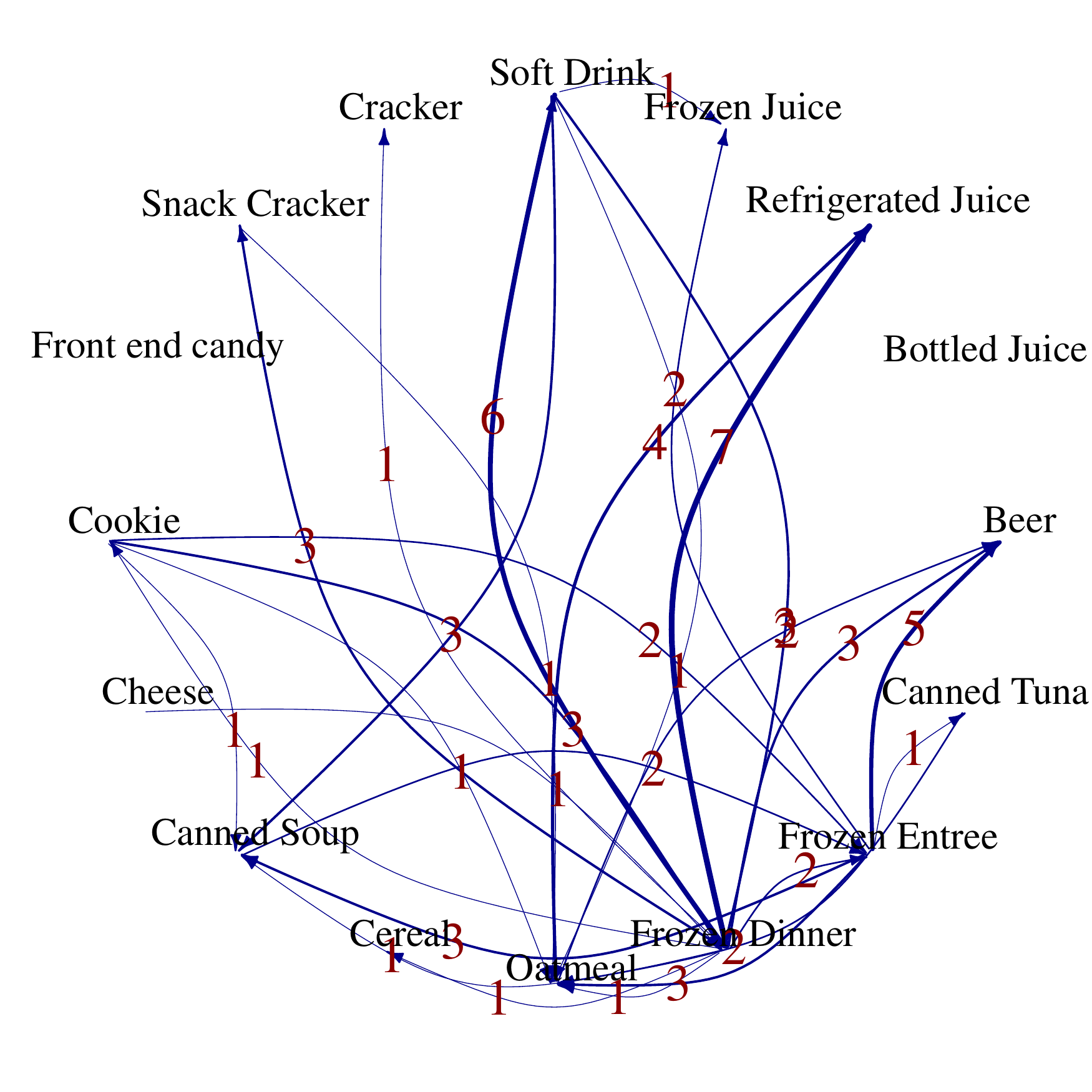}
        \caption{Network graph}
        \label{fig:3wayc}
    \end{subfigure}
    \caption{Monthly sales growth (log difference of sales) for $d = 16$ product categories from a single DominickÕs store (a U.S. supermarket chain; store \#21) with series length  $T = 76$, with observations from January 1993 to July 1994.}\label{fig:3way}
\end{figure}

We now consider an illustrative example using a marketing application. 
We consider the monthly sales growth (log difference of sales) for $d = 16$ product categories from a single DominickÕs store (a U.S. supermarket chain; store \#21) with series length  $T = 76$, with observations from January 1993 to July 1994. The product categories include Beer, Bottled Juices, Refrigerated Juices, Frozen Juices, Soft Drinks, Crackers, Snack Crackers, Front end candies, Cookies, Cheeses, Canned Soup, Cereals, Oatmeal, Frozen Dinners, Frozen Entrees and Canned Tuna. 
A panel of the 16 time series is shown in Figure \ref{fig:3waya}; each of the individual time series appears approximately stationary, sales growth goes not appear to have any clear time trends. For more information see \citet{Srinivasan04}, \citet{Gelper16}, and \url{https://research.chicagobooth.edu/kilts/marketing-databases/dominicks}.

The componentwise, elementwise, and own/other hierarchical lag-based penalty structures were fit to the weekly sales growth data.
The maximal VAR model orders were each specified as $p=10$.
Following \citet{nicholson2014hierarchical}, an expanding window was used in cross-validation to select penalty parameters for each method using the minimum one step ahead mean square forecast error (MSFE) criterion, initialized from the first third of the observations and running through the second third. Given the best penalty parameter value from validation, the MSFE was then computed out of sample for the final third of the observations. The MSFE for the HVAR$^{C}$, HVAR$^{O}$, HVAR$^{E}$ methods were 
0.970, 0.964, and 0.950, respectively. Although the HVAR$^{E}$ method had the lowest MSFE over this evaluation period, the approximate standard error for each was about 0.091. For comparison, forecasts based on lasso-VAR($p=10$), the running sample mean and from the random walk model (previous observation) were 1.120, 1.174, and 3.785, respectively. Although there was not much separation among the HVAR methods, they all performed significantly better than these competing methods. 

A graphical summary for the fitted HVAR$^{E}$ model is also provided using additional functions from the {\tt bigtime} package. 
The estimated lag matrix is shown in Figure \ref{fig:3wayb}. We note that most of the elements are estimated as zero (white), nearly all diagonal elements are active (1 or larger), the largest active lag index is 7, while most are 1 or 2. The most interesting lag-lead structure appears to be related to Frozen Dinners and Frozen Entrees, but we will forgo further interpretations. Another visualization of the lag-lead relationships between the variables using a directed, edge labeled network graph is shown in Figure \ref{fig:3wayc}. Here, Frozen Dinners and Frozen Entrees have the highest connectivity, while the overall network is relatively sparse.

\section{Conclusions}
Recent methodological and theoretical advances in stochastic regression and vector autoregression in high-dimension have provided key insights into the fundamental statistical issues of estimating high-dimensional models with dependent data. However, many interesting questions remain open from both theoretical and modeling  standpoint. From a theoretical perspective, principled selection of tuning parameters that explicitly take into account dependence information is an important direction. Also, estimating standard errors and quantifying uncertainty of sparse estimates are of particular interest. Within the frameworks of regression and VAR, developing asymptotic theory for sparse estimation of nonlinear, long memory and nonstationary processes with possibly heavy-tailed distributions are natural next steps. Some recent works have made progress in developing VAR type systems identification methods for generalized linear models and discrete-valued time series \citep{mark2018network, chen2017multivariate, hall2016inference}. From a modeling perspective, developing estimation methods for more general time series models like VARMA, ARCH, GARCH or cointegrated processes will have important applications in economics and finance. Incorporating additional structural information about the time series have also been effective in reducing dimensionality \citep{schweinberger2017high, melnyk2016estimating}. Another stimulating research direction is developing graphical modeling frameworks for time course data. Existing methods have primarily focused on precision matrix estimation, but this approach fails to capture the full conditional independence relationships among multiple time series by ignoring lagged dependencies. Frequency domain modeling, including estimation of spectral density and its inverse, can potentially overcome these limitations. Developing rigorous and efficient estimation methods for these problems will require developing novel probabilistic results for high-dimensional dependent random variables, as well as efficient computational algorithms, providing a promising research direction bringing together ideas from both theoretical and computational aspects of statistical research.

\section*{Acknowledgements}
We thank Dr. Ines Wilms for collection and analysis of the marketing data example. SB was supported in part by NSF award DMS-1812128, and NIH awards R01GM135926 and R21NS120227. DSM was supported by NSF (1455172, 1934985, 1940124, 1940276), Xerox PARC, the Cornell University Atkinson Center for a Sustainable Future (AVF-2017), USAID, and the Cornell University Institute of Biotechnology \& NYSTAR.

%\bibliography{references.bib}
\bibliographystyle{abbrvnat}
\bibliography{biblio-wires}  %,varxlbib,harbib,biblio

% \subsection*{\sffamily \Large FURTHER READING}
% Please insert any further reading/resources here.
% For readers who may want more information on concepts in your article, provide full references and/or links to additional recommended resources (books, articles, websites, videos, datasets, etc.) that are not included in the reference section. Please do not include links to non-academic sites, such as Wikipedia, or to impermanent websites.

\end{document}